\documentclass[aps,prl,amsmath,amssymb,prereprint,superscriptaddress]{revtex4-1}
\usepackage{graphicx}
\usepackage{amsmath}
\usepackage{dcolumn}
\usepackage{bm}
\usepackage{bbold}
\usepackage{color}
\usepackage{ulem}
\usepackage{nicefrac}

\bibstyle{apsrev4-1}

\begin{document}

\title{Observation of Rayleigh phonon scattering through excitation of extremely high overtones in low-loss cryogenic acoustic cavities for hybrid quantum systems}

\author{Maxim Goryachev}
\affiliation{ARC Centre of Excellence for Engineered Quantum Systems, University of Western Australia, 35 Stirling Highway, Crawley WA 6009, Australia}

\author{Daniel L. Creedon}
\affiliation{ARC Centre of Excellence for Engineered Quantum Systems, University of Western Australia, 35 Stirling Highway, Crawley WA 6009, Australia}

\author{Serge Galliou}
\affiliation{Department of Time and Frequency, FEMTO-ST Institute, ENSMM, 26 Chemin de l'\'{E}pitaphe, 25000, Besan\c{c}on, France}

\author{Michael E. Tobar}
\email{michael.tobar@uwa.edu.au}
\affiliation{ARC Centre of Excellence for Engineered Quantum Systems, University of Western Australia, 35 Stirling Highway, Crawley WA 6009, Australia}

\date{\today}


\begin{abstract}
The confinement of high frequency phonons approaching 1 GHz is demonstrated in phonon-trapping acoustic cavities at cryogenic temperatures using a low-coupled network approach. The frequency range is extended by nearly an order of magnitude, with excitation at greater than the 200$^{\text{th}}$ overtone achieved for the first time. Such high frequency operation reveals Rayleigh-type phonon scattering losses due to highly diluted lattice impurities and corresponding glass-like behaviour, with a maximum $Q_L\times f$ product of $8.6\times 10^{17}$ at $3.8$K and $4\times10^{17}$ at $15$mK. This suggests a limit on the $Q\times f$ product due to unavoidable crystal disorder. Operation at 15 mK is high enough in frequency that the average phonon occupation number is less than unity, with a loaded quality factor above half a billion. This work represents significant progress towards the utilisation of such acoustic cavities for hybrid quantum systems. 

\end{abstract}

\maketitle

The equilibrium ground state of hybrid mechanical systems has become a topic of interest of many research groups\cite{Schliesser2010}. To achieve operation in the equilibrium ground state, low temperatures and high frequencies are required to comply with the well-known rule $\hbar\omega \gg k_B T$. However, the operating frequency is limited to only a few megahertz for most high quality mechanical and acoustic systems such as toroids, cantilevers, and tuning forks. High frequency acoustic devices like thin-film Bulk Acoustic Resonators (FBAR) and high-overtone Bulk Acoustic Resonators (HBAR) exist and operate at frequencies up to several gigahertz, but at the cost of high losses. For instance, O'Connell et al.\cite{OConnell:2010fk} demonstrated single phonon control of an FBAR resonator at $6.07$~GHz, but exhibiting a quality factor of only $260$. This low value of $Q$ significantly limits the coherence time of hybrid systems based on these devices. 

The main limitation on quality factor in acoustic devices based on films is lack of phonon trapping. The inability to provide effective trapping results in the loss of phonons into the substrate and clamping mechanism, and is the reason why the losses of such devices do not drop significantly between ambient and liquid helium temperature. In contrast, Bulk Acoustic Wave (BAW) technology provides such effective trapping of energy carriers that acoustic losses are mostly limited by the material itself\cite{galliou:091911,Galliou:2008ve,NatSR}. As a result, the losses in these devices experience a significant drop of several orders of magnitude between ambient and liquid helium temperature. BAW cavities can be understood as acoustic analogues of the Fabry-P\'{e}rot cavities widely used in modern experimental physics. Devices employing Surface Acoustic Waves (SAW) are another candidate that can benefit from phonon trapping techniques, and are already used in some experiments\cite{Carmon1,Carmon2}, however the $Q\times f$ products of these devices are still orders of magnitude lower than that of BAW technology. 

In previous investigations, the operating frequency of the BAW cavities utilised was limited to around $100$~MHz. Such frequencies are not high enough to bring the device into its quantum ground state using traditional means of cooling such as a dilution refrigerator. Typical temperatures of $10-20$~mK achievable in such a system result in a thermal phonon occupation number of $n_{\text{TH}} \approx 10$ for such modes. Thus, additional techniques are required such as feedback and parametric cooling to enter the regime where $n_{\text{TH}} \ll 1$. 


It is important to underline the mechanisms that limit the resonant frequency of BAW devices. A natural consequence of the Landau-Rumer theory of acoustic losses\cite{landaurumer1,landaurumer2} is that quality factor $Q$ is independent of resonance frequency $f$, resulting in a law that $Q$ is a constant. This regime has been achieved at low temperatures and relatively high frequencies. At higher temperatures however, acoustic losses are mostly due to the Akhieser mechanism\cite{Akheiser} which asserts a $Q\times f = const.$ law. So, at room temperature, it is natural to think that higher operational frequencies impose limitations on the quality factor if the regimes of material losses is achieved (which is not the case for FBAR resonators). Thus, to achieve the highest quality factor in BAW devices, their frequencies are mostly limited artificially  to $5-10$ MHz. At the same time, increasing the overtone number results in excess loss and overdamping of the overtones. 

At cryogenic temperatures, near 4 K in particular, Landau-Rumer theory suggests that it is possible to increase the operating frequency of the device without the cost of increased losses. This can be done simply by exciting resonances with higher overtone (OT) numbers. However, in our previous investigations\cite{galliou:091911,Galliou:2008ve,Goryachev1}, no OTs greater than 23 were found, which limited resonance frequencies to $100$~MHz. This can be explained by an increase in losses due to scattering from resonator surface roughness and lack of effective phonon trapping.  
While the former can be remedied by better surface treatment, the latter can only be improved by better cavity design. For example, Stevens-Tiersten theory on the mode distribution in piezoelectric acoustic plates \cite{stevens:1811} uses linear elastodynamics as an approximation to crystalline solid phonon dynamics\cite{Maris1971}, and predicts the phonon probability distribution across the cavity plane. The theory predicts that phonon trapping can be improved by increasing the radius of curvature and size of the disk, which means that the out-of-cavity tails of the phonon probability distribution are reduced, such that the probability of phonons leaking into the environment is substantially decreased\cite{Aspelmeyer}. In principle, the trapping parameters $\alpha$ and $\beta$ are proportional to an OT number $n$. Therefore, phonon losses due to leakage through the suspension system should obey a $\nicefrac{Q}{F(n)}=const.$ law, where $F(n)$ is a polynomial function. 
In the present work, the role of cavity is played by a BAW quartz resonator (SC-cut\cite{1536996}, BVA type \cite{1537081}) with the fundamental frequency of the quasi-longitudinal mode $f_{\text{fun}}=3.138$~MHz. The unique feature of this device is the extremely large plate radius of $15$~mm which is more than twice as large as that used in our previous investigations\cite{Goryachev1}. This value is at the limit of current technological capabilities. Further details of the phonon trapping cavity design are given by Goryachev et al.\cite{Goryachev1}. Here, our investigations are solely concerned with longitudinally-polarized phonon modes, since they have demonstrated both theoretically\cite{landaurumer1,landaurumer2} and experimentally\cite{galliou:091911,Galliou:2008ve,Goryachev1} lower loss at low temperatures due to higher sound velocity. 

\begin{figure}[t!]
\centering
\includegraphics[width=3.5in]{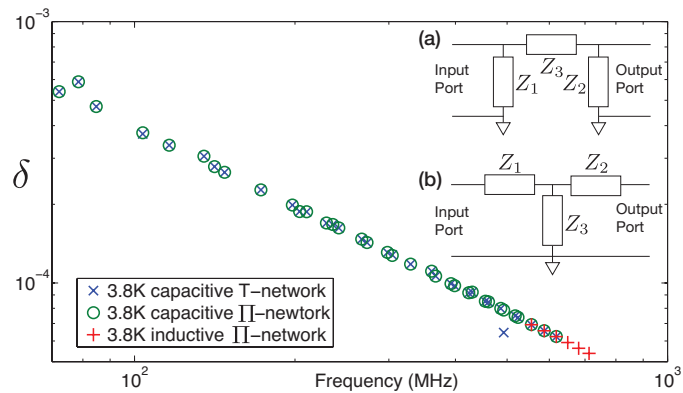}
\caption{\label{devi} Normalized deviation of the resonance frequency of overtones form the $n$th harmonic of the fundamental frequency $\delta = \frac{f_n-nf_{\text{fun}}}{nf_{\text{fun}}}$. The results are best approximated by $\delta\sim \nicefrac{1}{f}$ law. Inset: (a) Two-port $\Pi$ network topology, and (b) $T$ network topology.}
\end{figure}

BAW devices are one-port systems where the `port' is formed by two electrodes on opposite sides of the plate. Such devices are typically characterized with their impedance - a ratio between the complex amplitudes of voltage and current. This method is based on an impedance analyzer technique with appropriate calibration of the connecting cable\cite{Goryachev1}. The main disadvantage of this approach is its inability to decrease the driving power below the specifications of the impedance analyzer due to difficulties in separating the reflected wave at frequencies below 1 GHz. Thus, in the present experiment the BAW resonator is modified to become a two-port device. Since splitting the electrodes causes an unwanted perturbation to the trapping conditions, a more effective and simple way is to use the acoustic cavity as a building block for an electronic circuit.  
There are two topological ways to interconnect three electrical points (centres of input and output lines and the ground) using two-pin elements. In electrical engineering, they are known as $\Pi$ and $T$-circuits (see Fig.~\ref{devi} (a) and (b)). In order to decouple the acoustic cavity from both the source and sink, and reduce internal dissipation, $Z_3$ is taken to be the cavity itself, and $Z_1$ and $Z_2$ are capacitors (capacitive coupling) or inductors (inductive coupling). 
Reactive impedance of network elements decreases acoustic cavity coupling to the environment and thus the external dissipation.
Thus, although the measured quality factor must be considered as loaded by external circuit elements, the measurements do give reliable order-of-magnitude estimates. 
The BAW cavity-based networks are characterised using a transmission method. Three possible combinations (capacitive $\Pi$, capacitive $T$ and inductive $\Pi$ networks) were tested 
at $4$K. Under these conditions, many high quality modes demonstrated significant nonlinearity for typical values of the network analyzer output power.
Thus, the injected signal was attenuated at room temperature to give a total incident power of $0.1-0.3$ nW. 


Using this technique, extremely high overtones may be excited and characterized: up to the 197$^{\text{th}}$ overtone ($618.305$~MHz) for capacitive networks, and up to the 227$^{\text{th}}$ overtone ($712.456$~MHz) for inductively coupled networks. It must be mentioned that such high overtones have never been observed in phonon-trapping BAW devices before. The corresponding wavelengths are approximately $10$ and $8$~$\mu$m, which are comparable to the size of the gap between the plate and the electrodes.
It should be noted that due to boundary conditions of the system and the nature of piezoelectricity, only odd wave number overtones (corresponding to anti-symmetric types of vibrations) can be excited piezoelectrically in such devices. Amongst these overtones, only certain ones could be excited experimentally. They are resonances whose overtone number $n$ ends with $7$ (16 times), $5$ (14 times) and $3$ (5 times) giving clear periodicity (close to $10n$ with distribution $\pm 2n$) in trapping conditions.

The resonant frequencies of plate vibrations are solutions of the transcendental equation that can be derived from the partial differential equations of motion, and appropriate electrical and mechanical boundary conditions\cite{Tiersten1969,Tiersten1970}. Consequently, resonance frequencies are not exact integer multiples of the fundamental mode frequency (i.e. harmonics). The theory predicts that the deviation of an actual resonance frequency $f_n$ from the integer multiple of the fundamental frequency $n\times f_{\text{fun}}$ decreases with overtone number. Our experimental results confirm this prediction and follow a power law (see Fig.~\ref{devi}). The results show that for all three types of network studied, the frequencies measured for overtones of the same number are very similar. Except for a few overtones, the typical normalised frequency difference between capacitively coupled networks is of the order of $10^{-7}$. This confirms the common nature of the resonances characterized which should be attributed to the acoustic cavity, with only minimal influence from the external circuit. 
 
\begin{figure}[t!]
\centering
\includegraphics[width=3.5in]{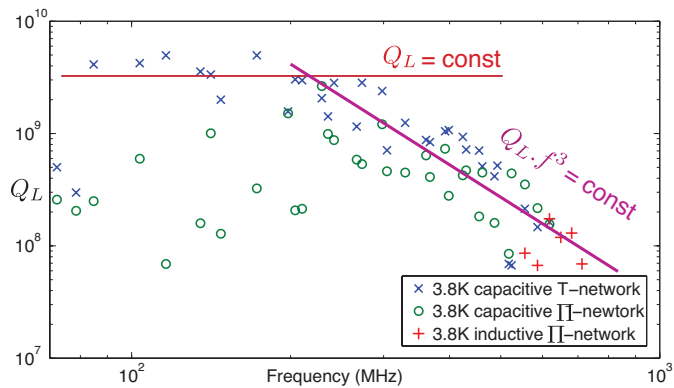}
\caption{\label{Q4K} Loaded quality factors estimated for different OTs  for different networks at $3.8$~K. Note that low frequency results (below $200$ MHz) specially for the $\Pi$-network are limited by losses in the load.}
\end{figure}

The estimated values of quality factor for all three types of networks are given in Fig.~\ref{Q4K}. Although these values are measured without calibration, and thus depend on the excitation and measurement circuits, the results are a good estimation of cavity acoustic losses. Moreover, for high resonance frequencies the overtones are only very weakly coupled to the environment, which means that $Q_L$ approaches the intrinsic quality factor $Q_0$ as the resonant frequency increases. 
The maximum value of loaded quality factor ($\sim$5 billion) was achieved for the 37$^{\text{th}}$ ($116.160$~MHz) and 55$^{\text{th}}$ ($172.651$~MHz) OTs for the capacitive $T$-network. 13 OTs exhibit quality factors above 1 billion with corresponding linewidths on the order of a few millihertz. The highest frequency with a loaded quality factor over 1 billion is $398.616$~MHz (127$^{\text{th}}$ OT), while the capacitive $\Pi$-network exhibits higher losses due to its stronger coupling to the environment. The highest $Q_L\times f$ product achieved at this temperature is $8.6\times 10^{17}$ at $172.7$ MHz. 

Each network demonstrates similar dependence of the quality factor on frequency. The maximum quality factor achieved was in the frequency range of $100-200$~MHz. The $Q=const.$ dependence is clearly observed for the capacitive $T$-network, which demonstrates $Q_L\rightarrow Q_0$. Above these frequencies, the quality factor drops significantly according to a $Q_L\times f^3 = const$ law. This law is independent of the type of network used, and is in the region of frequency where acoustic modes are very weakly coupled to the electrical circuit, so the losses can be attributed to the acoustical cavity itself. This type of $Q(f)$-dependence corresponds to elastic Rayleigh scattering of waves over microscopic particles whose dimensions are smaller than a wavelength. This type of wave scattering is the main mechanism of loss in optical fibers\cite{Zhi}. Rayleigh scattering is characterised by attenuation inversely proportional to the forth power of the wavelength, or equivalently directly proportional to the fourth power of frequency: $\alpha\sim f^4$. So, using the relation $Q\sim\nicefrac{f}{\alpha}$ gives the observed $Q\sim\nicefrac{1}{f^3}$ law. This type of scattering originates from microscopic impurities randomly distributed in the quartz crystal, such as the impurity ions Al$^{3+}$, Na$^+$, Li$^+$, Se$^{4+}$, and H$^+$~ \cite{Martin, MasonWP, Euler}. The acoustic cavity was manufactured from highest quality alpha synthetic quartz, and has undergone special purification treatment, however the concentration of these impurities can still be as high as $1-3$ parts per billion. Other sources of disorder in the crystal can be attributed to phenomena such as dislocations and microscopic structural defects.
Acoustic wave scattering over this quenched randomness is analogous to a light beam scattering in optical fibers. The mechanism can be understood by a direct analogy with the corresponding optical system - the Fabry-P\'{e}rot cavity. The longitudinally polarised phonons in a plate reflected by  surfaces are analogous to photons in a Fabry-P\'{e}rot cavity reflected by mirrors. In such structures, resonant \hbox{(quasi-)particles} have a certain direction of propagation normal to the mirrored surfaces. 
When a phonon hits an impurity or other imperfection of the crystal structure, it can maintain the same direction of propagation without attenuation, or alternatively change the direction of propagation and escape the cavity\cite{Aspelmeyer}, which results in energy loss.  Rayleigh scattering of acoustic waves has been theoretically predicted and observed for various polycrystalline and glass materials~\cite{mason47,bhatia58,Papadakis65}, i.e. disordered solids. In the present work, we observe this type of scattering in the form of losses in an extremely high purity crystalline structure due to exceptional sensitivity. Thus, a crystal with dilute impurities behaves as crystalline solid for lower order OTs and demonstrate glass-like behaviour at higher frequencies. 


Another mechanism limiting the excitation of higher overtones is related to the shunt capacitance formed by the electrodes of the device. This capacitance short-circuits the vibrating plate at higher frequencies, decoupling it from the outside world. Inductive coupling partially compensates this capacitance leading to excitation of even higher overtones.


In a separate experiment, the capacitive $T$ network was cooled down to millikelvin temperatures using a cryogen-free dilution refrigerator system from BlueFors Cryogenics. The device temperature was stabilised at about $15$~mK. In order to avoid all nonlinear and thermal effects, the circuit incident power was cryogenically attenuated by 40 dB, preventing injection of room-temperature thermal noise, and by 30 dB at room temperature, leading to a minimum network incident power of $-115$~dBm. The transmitted output signal was amplified cryogenically ($4$K) and at room temperature.

\begin{figure}[t!]
\centering
\includegraphics[width=3.4in]{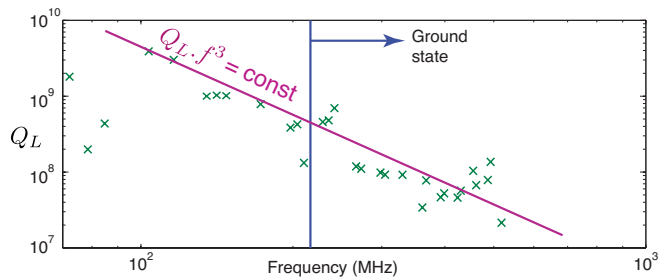}
\caption{\label{QmK} Loaded quality factors estimated for different OTs for the capacitive $T$-network at $15$~mK. For frequencies below $100$~MHz, losses are due to external load.}
\end{figure}

The normalized resonant frequency deviation at this temperature was the same as shown in Fig.~\ref{devi}. For all overtones, the resonance frequencies decreased by $37-100$~Hz between $3.8$~K and $15$~mK. 
The distribution of quality factors at millikelvin temperature for different overtones is shown in Fig.~\ref{QmK}. The values of quality factor for OTs above $110$~MHz at this temperature are lower than at liquid helium temperatures (Fig.~\ref{Q4K}). Our results (see Fig.~\ref{QmK}) demonstrate that this is due to excess Rayleigh scattering losses dominating over the Landau-Rumer losses due to scattering of phonons by thermal phonons over the crystal lattice anharmonicity. This increase of the observed losses with decreasing temperature is also in accordance with the Rayleigh scattering mechanism.  More scatterers are found in their ground states at lower temperatures, meaning that more two-level systems can participate in the process by absorbing acoustic phonons propagating in the direction normal to the plate, and re-emitting them in a random direction. 
Nevertheless, the highest achieved $Q_L\times f$ product of $4.0\times10^{17}$ exceeds our previous results of intrinsic $Q\times f = 7.8\times10^{16}$. Additionally, in the range of $150-300$~MHz, $Q_L$ drops faster with frequency. 

According to the Bose-Einstein distribution, in order to achieve an average value of thermal phonons less than unity at $15$~mK, the resonance frequency must exceed $216.65$~MHz. This criterion is satisfied by overtones equal to and higher than the $73^{\text{rd}}$ OT (19 in total). The lowest average thermal occupation number is $0.2$ for the highest overtone characterised at this temperature (165$^{\text{th}}$ OT with a frequency of $517.875$~MHz). The maximum quality factor measured for these OTs is $696$ million.


One of the goals of the present study is to improve the sensitivity in characterisation of acoustic devices at millikelvin temperatures. For this purpose, a cascade of cold attenuation on the input line and cold amplification on the output line was built. 
This allowed characterization of the power dependence of the OTs down to $10^{-12}$~W (parasitic attenuation not included) corresponding to $10^6-10^7$ stored acoustic phonons.

\begin{figure}[t!]
\centering
\includegraphics[width=3.5in]{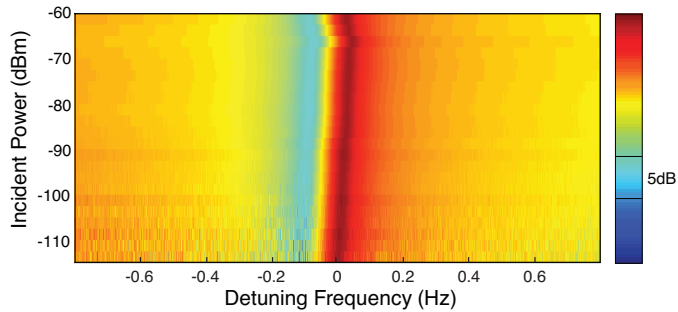}
\caption{\label{nonlin} Power dependence of the magnitude of the network transmission coefficient.}
 \end{figure} 

Two of the OTs (33$^{\text{rd}}$ and 37$^{\text{th}}$) demonstrated significant power dependence even at the lowest achievable power. Figure~\ref{nonlin} shows that the resonance frequency of the 37$^{\text{th}}$ OT linearly changes with decreasing power. Nevertheless, the resonance bandwidth does not demonstrate any significant change. This situation could be modelled with a Kerr-like term in the corresponding equations of motion of linear elasticity. This type of nonlinearity originates in the anharmonicity of the crystal lattice which gives birth to four acoustic phonon mixing terms. The existence of a turn-over point near $-65$~dBm demonstrates that the nonlinearity is higher than third order (in the equations of motions). 
Such high power sensitivity is due to the extremely low losses of these OTs.

Low loss phonon-trapping acoustic cavities demonstrate high potential for applications to quantum hybrid systems,  and as a test-bench for various physical experiments including tests of fundamental physics~\cite{aspelmayer2,NatSR}. Besides extremely low losses at relatively high frequencies, these devices are easy to handle and integrate with external electronic devices to build complex networks.  
We have demonstrated experimentally that improved trapping in such devices can lead to significant improvement in $Q\times f$ products. The fundamental limitations of the $Q\times f$ product is related to the existence of highly dilute impurities which results in Rayleigh scattering losses. Thus, the same material exhibits both crystalline (Landau-Rumer) and glass-like (Rayleigh scattering) features at different frequencies. This limits the ultimate values of the $Q\times f$ product due to the inability of current technology to manufacture ideally pure crystals.
 
\section*{Acknowledgements}
This work was supported by the Australian Research Council Grant No. CE110001013 and FL0992016, and the Regional Council of Franche Comt\'{e}, France, Grant No. 2008C16215.

%

\end{document}